\documentclass[prl, showpacs, twocolumn, floatfix]{revtex4}

\usepackage{graphicx}
\usepackage{amsmath, amsfonts, amssymb, bm}
\usepackage[]{psfrag}
\newcommand{\dbar}{\raisebox{0.44em}
                  {\rule{1ex}{0.04em}}\hspace{-1.2ex}\lambda_C}

\psfragscanoff
\begin{document}

\title{Muon pair creation by two x-ray laser photons in the field of an atomic nucleus}

\author{Carsten M\"uller}
%\email{c.mueller@mpi-hd.mpg.de}
\author{Carlus Deneke}
%\email{c.mueller@mpi-hd.mpg.de}
\author{Christoph H. Keitel}
%\email{keitel@mpi-hd.mpg.de}
\affiliation{Max-Planck-Institut f\"ur Kernphysik, Saupfercheckweg 1, 69117 Heidelberg, Germany}

\date{\today}

\begin{abstract}
The generation of muon-antimuon pairs is calculated in the collision of an ultrarelativistic bare ion with an intense x-ray laser beam. The reaction proceeds nonlinearly via absorption of two laser photons. By systematic study throughout the nuclear chart we show that the interplay between the nuclear charge and size, along with the possibility of nuclear excitation leads to saturation of the total production rates for high-$Z$ ions, in contrast to the usual $Z^2$ scaling for pointlike projectiles. The process is experimentally accessible by combining present-day ion accelerators with near-future laser sources and in principle allows for the measurement of nuclear form factors.
\end{abstract}
 

\maketitle

In recent years the interest in electron-positron pair creation in combined laser and nuclear Coulomb fields has been stimulated by the large advances in high-power laser devices and ion accelerators (see \cite{e+e-1,e+e-2,Kuchiev} and references therein). The process is interesting since it proceeds by the absorption of real photons from the laser field and a virtual photon from the Coulomb field. Most studies have considered the low-frequency tunneling regime, where laser field strengths close to the Schwinger value $E_S^{(e)}\equiv m_e^2c^3/\hbar e=1.3\times 10^{16}\,{\rm V/cm}$ are required. At this field intensity, the electric work performed on an electron of charge $e$ along the Compton wavelength $\dbar^{(e)}=\hbar/m_ec$ equals its rest energy $m_ec^2$. In the opposite high-frequency multiphoton regime, pairs are created via the simultaneous absorption of several high-energy photons with $\hbar\omega\sim m_ec^2$ from a rather weak laser field. In both cases, the presence of the nucleus is required to guarantee energy-momentum conservation.

The field strengths and frequencies required for $e^+e^-$ pair creation in the respective regimes are four orders of magnitude larger than provided by the most-advanced present laser systems.
This gap can be bridged in laser-ion collisions, where the ions are counterpropagating the laser beam at relativistic speed \cite{e+e-1,e+e-2,Kuchiev}. Then the laser field strength and frequency as seen by the nucleus in its rest frame are enhanced by the relativistic Doppler shift. The only experimental observation of $e^+e^-$ pair production in laser and Coulomb fields has been achieved at SLAC (Stanford, California) in collisions of ultrarelativistic electrons with an intense optical laser pulse \cite{SLAC}. The pairs were generated indirectly via Compton backscattering and a subsequent collision of real photons.

In view of the ongoing technical progress the question arises, whether not only $e^+e^-$ but also $\mu^+\mu^-$ pairs can be produced with the emerging near-future laser sources. Muon production in the tunneling regime appears rather hopeless, though, since the required field needs to be close to $E_S^{(\mu)}=\rho^2E_S^{(e)}=5.6\times 10^{20}\,{\rm V/cm}$, with the muon-to-electron mass ratio $\rho=m_\mu/m_e\approx 207$. Even by exploiting the $\gamma$-factors which will soon be provided by the Large-Hadron Collider  at CERN (Geneva, Switzerland) this value seems out of reach ($\gamma=7000$ for protons \cite{LHC}). As an alternative, it has recently been proposed that $\mu^+\mu^-$ pairs could be produced in laser and Coulomb fields within a two-step process \cite{Kuchiev}: First, an $e^+e^-$ pair is created via tunneling in a laser-ion collision, and afterwards driven by the laser field into an annihilating recollision $e^+e^-\to\mu^+\mu^-$. The latter process has also been studied in \cite{PRD}, with the electron and positron originating from a positronium atom or an $e^+e^-$ plasma.

In this Letter, we calculate direct $\mu^+\mu^-$ production in highly energetic laser-ion collisions. Rather than the tunneling regime, we consider multiphoton muon production via absorption of few high-frequency laser photons. The latter are assumed to stem from an x-ray free electron laser (XFEL) beam with $\hbar\omega_0=12$\,keV \cite{XFEL}, which collides head-on with an ultrarelativistic nucleus moving at $\gamma = 7000$. In the nuclear rest frame the photon energy amounts to $\hbar\omega\approx 2\gamma \hbar\omega_0 = 168$\,MeV, so that the energy threshold $\Delta\epsilon = 2m_\mu c^2$ for $\mu^+\mu^-$ production can be overcome by two-photon absorption from the XFEL field \cite{e-impact}. Corresponding large-scale XFEL facilities are presently being developed at SLAC and DESY (Hamburg, Germany), where peak intensities close to 10$^{20}$\,W/cm$^2$ are envisaged \cite{XFEL}. The projectile nucleus is modelled by an extended charge distribution, whose shape is shown to have significant impact on the muon creation rates.

At first sight, $e^+e^-$ and $\mu^+\mu^-$ pair production in combined laser and Coulomb fields seem to be very similar processes since the electron and muon only differ by their mass (and lifetime). In this picture, the corresponding production probabilities would coincide when the laser field strength and frequency are scaled in accordance with the mass ratio, i.e. $W_{\mu^+\mu^-}(E^{(\mu)},\omega^{(\mu)})=W_{e^+e^-}(E^{(e)},\omega^{(e)})$ for $E^{(\mu)}=\rho^2E^{(e)}$ and $\omega^{(\mu)}=\rho\omega^{(e)}$. This simple scaling argument does not apply, however, as the large muon mass is connected with a correspondingly small Compton wavelength $\dbar^{(\mu)}=\dbar^{(e)}/\rho\approx 1.86\,$fm, which is smaller than the radius of most nuclei. As a result, the nucleus does not look pointlike to the muon and its finite extension must be taken into account.  Pronounced nuclear size effects have also been found for $\mu^+\mu^-$ production by single $\gamma$-photon impact on nuclei \cite{Tsai} and in relativistic heavy-ion collisions \cite{BeBa,two-photon}. 

Following the usual theoretical approach to lepton pair creation in combined laser and Coulomb fields (see, e.g., \cite{e+e-2}), we write the process amplitude in the nuclear rest frame as
\begin{eqnarray}
\label{S}
S_{p_+p_-} = -\frac{ie}{\hbar}\int dt\int d^3r\,\Psi_{p_-}^\dagger V(r)\Psi_{p_+}.
\end{eqnarray}
The muons are created with free momenta ${\bf p}_\pm$ and described by relativistic Volkov states $\Psi_{p_\pm}$ \cite{LL} which include their interaction with the laser field to all orders. The nuclear field $V(r)$ is taken into account within the first-order of perturbation theory. We assume the nucleus to be spherically symmetric with a Gaussian charge distribution
\begin{eqnarray}
\label{rho}
\varrho(r) = \frac{Ze}{(\sqrt{\pi}a)^3}\,{\rm e}^{-r^2/a^2},
\end{eqnarray}
where $Z$ is the atomic number and the parameter $a$ is related to the nuclear rms charge radius by $r_{\rm rms}=\sqrt{3/2}a$. The charge density \eqref{rho} will serve us for a systematic study of the nuclear size effect on the muon creation process. We note that light nuclei with mass number $A\sim 10$ are Gaussian-shaped to a good approximation, whereas the charge density of heavy isotopes ($A\sim 100$) is usually parametrized by a Fermi distribution \cite{Povh}. The electrostatic potential generated by the charge density (\ref{rho}) is $V(r)=(Ze/r)\Phi(r/a)$, with the error function $\Phi(x)$. The amplitude in Eq.\,(\ref{S}) can be evaluated analytically by expanding its periodic part into a Fourier series. This way, one arrives at the integral
\begin{eqnarray}
\label{Vq}
\tilde V(q) = \int d^3r\, V(r)\,{\rm e}^{\rm \frac{i}{\hbar}\bf{qr}} 
= \frac{Ze}{q^2}\,{\rm e}^{-(qa/2\hbar)^2},
\end{eqnarray}
where ${\bf q}\equiv \tilde{\bf p}_+ + \tilde{\bf p}_- - n{\bf k}$ is the momentum transfer to the nucleus. Within the multiphoton regime, the laser-dressed momenta $\tilde {\bf p}_\pm$ introduced here practically coincide with the free momenta ${\bf p}_\pm$ \cite{LL}. Moreover, $n=2$ laser photons of momentum ${\bf k}$ are absorbed from the field. The first factor on the right-hand side of Eq.\,(\ref{Vq}) is the Fourier transform of the Coulomb potential of a pointlike nucleus. The second factor is the elastic nuclear form factor $F(q^2)$, which describes the correction due to the finite nuclear size and leads to substantial reduction of the process probability when $q\sim m_\mu c$ approaches or exceeds the inverse nuclear radius. The fully differential rate for two-photon muon pair creation is obtained by summing the square of the amplitude~(\ref{S}) over the lepton spins:
\begin{eqnarray}
\label{R}
dR_{\rm el} = \sum_{\rm spins} |S_{p_+p_-}|^2 \frac{d^3p_+}{(2\pi\hbar)^3} \frac{d^3p_-}{(2\pi\hbar)^3}.
\end{eqnarray}
Equation~(\ref{R}) accounts for the elastic channel of the process, where the nucleus remains in its ground state. It adopts the structure $dR_{\rm el}=dR_0 Z^2 F^2(q^2)$, with $dR_0$ denoting the differential production rate for a pointlike proton. This structure is analogous, e.g., to the form factor-corrected Rutherford cross section or the Rosenbluth formula for electron-nucleon scattering, which also factorize into a product of the result for a point scatterer and the form factor squared \cite{Povh}. An additional contribution to $\mu^+\mu^-$ creation comes from the inelastic channel, where the nucleus is excited due to the recoil imparted on it \cite{Tsai,Thomas}. The $Z$ protons inside the nucleus act incoherently here, and one has to a good approximation $dR_{\rm inel}\approx dR_0 Z (1-F^2(q^2))$. The inelastic process becomes important when $F^2\ll 1$. In principle, both production channels are separately accessible in experiment by observing the final nuclear state in coincidence.

\begin{figure}[b]
\begin{center}
%\resizebox{8cm}{!}{\includegraphics{angleRel.eps}}
\resizebox{8cm}{!}{\includegraphics{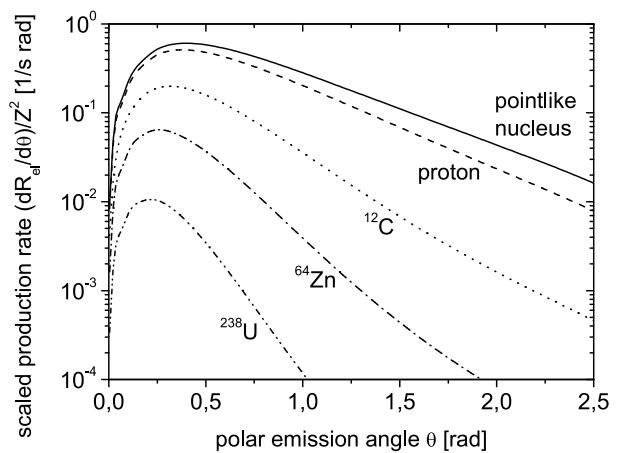}}
\caption{Angular spectra of one of the muons produced by two-photon absorption from
an intense XFEL beam ($\hbar\omega_0=12$\,keV, $\xi_\mu=6.8\times 10^{-5}$) colliding with
various ultrarelativistic nuclei ($\gamma=7000$), as indicated. The rates are taken in the nuclear rest frame and scaled by the nuclear charge.} 
\end{center} 
\end{figure}

Figure~1 demonstrates the nuclear size effect by showing angle-differential rates for elastic muon pair creation in ultrarelativistic collisions of various ions with an intense XFEL beam of linear polarization. The emission angle refers to the laser wave vector and the rates are scaled by $1/Z^2$. The laser intensity parameter $\xi_\mu=\frac{eE}{m_\mu c\omega_0}$ relates to the muon mass; the value chosen corresponds to $2.54\times 10^{22}$\,W/cm$^2$. We point out that this intensity exceeds the original design values at SLAC and DESY by 2-3 orders of magnitude, but can be attained by anticipated facility extensions and improved x-ray focussing techniques \cite{futureLCLS}. The respective rms charge radii are 0.875\,fm (proton), 2.470\,fm ($^{12}$C), 3.929\,fm ($^{64}$Zn), and 5.851\,fm ($^{238}$U) \cite{radii}. For comparison we also show the result for a pointlike nucleus. 
With increasing nuclear size, the production rates are strongly reduced, and the maximum of the angular distributions is shifted towards smaller angles. The latter is because the nuclear form factor cuts the contributions from large momentum transfers, which otherwise give rise to large emission angles. The reduction of the total rates with respect to the result for a point charge arises from the fact that the particles are created at typical distances $\bar r\sim\dbar^{(\mu)}$. When this distance is smaller than the nuclear radius, only a fraction $eZ_{\rm eff} \equiv 4\pi\int_0^{\bar r} r^2\varrho(r)dr$ of the total nuclear charge contributes effectively to the process.

Figure~2 shows integrated rates of $\mu^+\mu^-$ production for several projectiles. The nuclear isotope of largest abundance was chosen always. The values again refer to the nuclear rest frame. For a point proton, the total rate would amount to $R_0=(\alpha^2/16\pi) \xi_\mu^4\,\omega\mathcal{F}(\omega)\approx 0.58$\,s$^{-1}$ (cp. Eq.~(26) in \cite{e+e-1}), displaying the typical $\xi_\mu^4$ dependence of a two-photon process. Here, $\alpha$ denotes the finestructure constant and the dimensionless function $\mathcal{F}(\omega)$ is of the order of 0.1 in the relevant frequency range (i.e., $1\le \hbar\omega/m_\mu c^2\le 2$).
As compared to the point-charge result, the elastic production rates are reduced by 23\% for proton impact and by 77\% for $^{12}$C impact. For the heavier projectiles, the reduction factors amount to 0.13 ($^{27}$Al), 0.087 ($^{40}$Ca), 0.055 ($^{64}$Zn), 0.042 ($^{84}$Kr), 0.026 ($^{120}$Sn), 0.020 ($^{142}$Nd), 0.010 ($^{208}$Pb), and $7.1\times 10^{-3}$ ($^{238}$U), respectively. Since the elastic process probability increases with the projectile charge as $Z^2$ and decreases with its size as $\sim\exp[-\frac{1}{3}(r_{\rm rms}/\dbar^{(\mu)})^2]$, a maximum elastic rate arises for atomic numbers around $Z\approx 60$. This is in contrast to $e^+e^-$ pair creation which increases as $Z^2$ throughout \cite{e+e-1,e+e-2,Kuchiev}, apart from higher-order corrections in $\alpha Z$ which slightly modify this behaviour at high $Z$. Note that the latter corrections are of minor importance in the present situation, as the muons are produced well above the energetic threshold. 
The emergence of the maximum in Fig.\,2 can be understood by consideration of the effective nuclear charge $Z_{\rm eff}$ contained in a sphere of radius $\dbar^{(\mu)}$. The corresponding values 3.66 ($^{84}$Kr), 3.83 ($^{120}$Sn), 3.96 ($^{142}$Nd), 3.96 ($^{208}$Pb), and 3.73 ($^{238}$U) agree with the observed location of the maximum. This simple picture also explains a local minimum for atomic numbers around $Z\approx 70$ (not shown), since $Z_{\rm eff}\approx 3.72$ for $^{174}$Yb. We stress that the maximum remains at the same position, when the nuclei are modelled by uniformly charged spheres, which represents a better approximation than Eq.~\eqref{rho} to the Fermi charge distribution of the relevant heavy isotopes where the maximum occurs \cite{Povh}. A maximum also arises for elastic muon production by a single photon of twice the energy.
Figure~2 moreover displays the total rate $R_{\rm tot}=R_{\rm el}+R_{\rm inel}$, which does not exhibit a maximum but saturates at high $Z$ values since the contribution from the inelastic channel increases with nuclear charge. The total rates are still considerably smaller than the point-proton results: The reduction factors amount to, e.g., 0.35 ($^{12}$C), 0.13 ($^{40}$Ca), 0.069 ($^{84}$Kr), 0.036 ($^{142}$Nd), and 0.018 ($^{238}$U). For $e^+e^-$ pair creation, the inelastic channel is negligible since $q\sim m_ec\ll \hbar/r_{\rm rms}$, so that $F^2\approx 1$ and $R_{\rm tot}\approx R_0Z^2$.

Figure~3 shows angular spectra in the laboratory frame for carbon impact. Apart from the Gaussian charge distribution of Eq.~\eqref{rho}, a uniform density within a sphere of radius $\sqrt{5/3}\,r_{\rm rms}$ was assumed, which has equal rms radius. As compared to the result for a Gaussian-shaped $^{12}$C nucleus, the elastic (total) production rate is smaller by 13\% (7\%) for the hard sphere. The relative difference grows for heavier nuclei, reaching about 40\% (20\%) for $^{238}$U. This demonstrates that the muon production process is sensitive not only to the nuclear radius, but also to the nuclear shape. In principle, the process might even be used for determination of the nuclear form factor, which is traditionally accomplished by electron scattering.

\begin{figure}[t]
\begin{center}
%\resizebox{8cm}{!}{\includegraphics{totalRinel.eps}}
\resizebox{8cm}{!}{\includegraphics{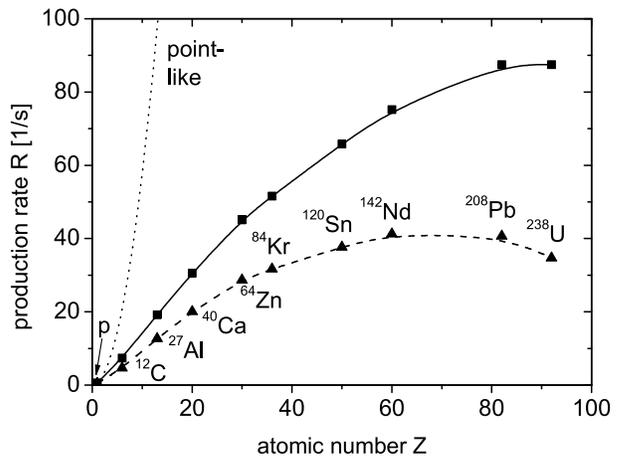}}
\caption{Nuclear-frame rates for muon creation in the XFEL-nucleus collisions as in Fig.\,1. The squares and triangles show the total and elastic rates for various nuclei, respectively, connected by fit curves. The dotted line $\propto Z^2$ holds for a point nucleus.}
\end{center} 
\end{figure}

\begin{figure}[t]
\begin{center}
%\resizebox{8cm}{!}{\includegraphics{angleC12inel.eps}}
\resizebox{8cm}{!}{\includegraphics{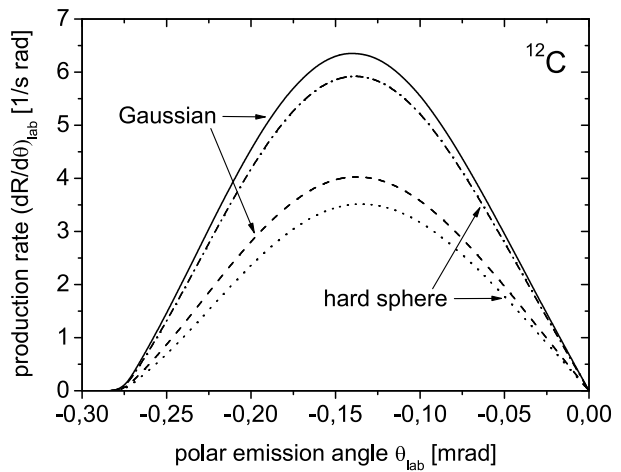}}
\caption{Angular spectra in the laboratory frame for two-photon muon pair production by  $^{12}$C nuclei of different shapes and an XFEL beam. The upper (lower) two curves show  total (elastic) rates. The collision parameters are as in Fig.\,1.}
\end{center} 
\end{figure}

According to our results, muon pair creation in XFEL-nucleus collisions is 
experimentally feasible by employing near-future technology. In the laboratory frame, the production rates of Fig.~2 are reduced by a factor $1/\gamma$ due to relativistic time dilation and, accordingly, reach values of about $R_{\rm lab}\approx 10^{-2}$\,s$^{-1}$. In the collision of an ion beam containing $N=10^{11}$ particles \cite{LHC} with an XFEL pulse of $\tau=100$\,fs duration \cite{XFEL}, the probability for production of one muon pair is $W_{\mu^+\mu^-}\approx R_{\rm lab}\tau N/2 \approx 10^{-4}$. At the envisaged average XFEL repetition rate of 40\,kHz, one muon production event per second is obtained. The muon yield scales with the laser intensity squared. The typical lab-frame energy of the muons $\sim 500$\,GeV is highly relativistic; the muonic lifetime is accordingly increased to $\sim 10$\,ms. $e^+e^-$ pairs are also produced in the collision by single-photon absorption in the nuclear field (Bethe-Heitler process \cite{LL}).
%with a rate of $\sim Z^2\times 10^{11}\,{\rm s}^{-1}$. 
We stress, however, that the subsequent reaction $e^+e^-\to\mu^+\mu^-$ via laser-driven recollisions \cite{Kuchiev} does not occur in the multiphoton regime considered here, because the relative electron-positron momentum $\Delta p\sim m_e c$ satisfies $0\ll\Delta p\ll 2m_\mu c$. Detected muons have thus been produced via the direct process $A+n\omega \to A + \mu^+\mu^-$, with $n=2$. Higher photon orders ($n\ge 3$) are suppressed by an additional factor $\xi_\mu^2\ll 1$.

In XFEL-nucleus collisions, also bound-free $\mu^+\mu^-$ pair production can take place, with the $\mu^-$ being created in a bound state of the atomic nucleus. In order to estimate the corresponding rate, we apply an appropriatly modified version of the theory developed in \cite{PRL} for $e^+e^-$ production with K-shell capture. Assuming the same collision parameters as in Fig.~2 and a pointlike proton, we obtain a rest-frame rate of $1.94\times 10^{-6}$\,s$^{-1}$. This value is by five orders of magnitude smaller than the corresponding rate for free muon pair creation in Fig.~2. For point projectiles, the bound-free production channel raises like $Z^5$ with atomic number \cite{PRL}. This scaling is also considerably damped by the finite nuclear size.

Muon pair creation has been studied before in relativistic heavy-ion collisions \cite{BeBa}. The process is formally related to muon production by a single photon in the field of an atomic nucleus \cite{Tsai,Sorensen} by the Weizs\"acker-Williams method, which describes the transverse electromagnetic field of an ultrarelativistic nucleus by a spectrum of virtual photons \cite{LL}. In accordance with the present results, the finite nuclear extent was found to reduce the production probability substantially. In the high-energy limit, the argument of the famous Bethe-Heitler logarithm $\log(2\hbar\omega/m_\mu c^2)$ is modified by the factor
$\dbar^{(\mu)}/r_{\rm rms}$ \cite{BeBa,Sorensen}. This factor is also responsible for the rate suppression of two-photon muon pair creation here. Numerical calculations of bound-free muon pair production in $^{238}$U-$^{238}$U collisions have found a cross section reduction by five orders of magnitude due to the nuclear extent \cite{Mombi}.
It is interesting to note that exclusive dilepton production in hadron collisions also represents a search tool for physics beyond the standard model \cite{exclusive}. 

In conclusion, direct production of $\mu^+\mu^-$ pairs by two-photon absorption from a high-frequency laser wave colliding with an atomic nucleus was calculated. The process is sensitive to the nuclear form factor. It could be realized experimentally by combining the radiation from upcoming XFEL sources with an ultrarelativistic ion beam from the present generation of heavy-ion accelerators. The highest elastic production rates are reached by projectiles in the lanthanoid region ($Z\approx 60$), whereas the total production rates saturate at high $Z$ values.

Useful input by T.~J.~B\"urvenich, K.~Z.~Hatsagortsyan, and A.~B.~Voitkiv is gratefully acknowledged.


\begin{thebibliography}{99}

\bibitem{e+e-1} 
A. I. Milstein {\it et al.}, Phys. Rev. A \textbf{73}, 062106 (2006).

\bibitem{e+e-2} 
J. Z. Kaminski, K. Krajewska, and F. Ehlotzky, Phys. Rev. A {\bf 74}, 033402 (2006).

\bibitem{Kuchiev} M. Yu. Kuchiev, Phys. Rev. Lett. \textbf{99}, 130404 (2007).

\bibitem{SLAC} D. Burke {\it et al.}, Phys. Rev. Lett. {\bf 79}, 1626 (1997).

\bibitem{LHC} W.-M. Yao {\it et al.}, J. Phys. G \textbf{33}, 1 (2006).

\bibitem{PRD} C. M\"uller, K. Z. Hatsagortsyan, and C. H. Keitel, Phys. Rev. D {\bf 74}, 074017 (2006); Phys. Lett. B {\bf 659}, 209 (2008).

\bibitem{XFEL} 
L. F. DiMauro {\it et al.}, J. Phys. Conf. Ser. \textbf{88}, 012058 (2007);
M.~Altarelli~{{\it et al.}}, Technical Design Report of the European XFEL, DESY 2006-097 (http://www.xfel.net).

\bibitem{e-impact} Due to recoil effects, the corresponding $\gamma$-factor of a projectile electron required for muon pair creation by two-photon absorption is much larger ($\gamma\gtrsim 10^6$).

%\bibitem{photon-hadron} A. K. {\c C}ift{\c c}i, S. Sultansoy, and {\"O}. Yava{\c s}, Nucl. Instrum. Meth. Phys. Res. A \textbf{472}, 72 (2001).

\bibitem{Tsai} 
A. Alberigi-Quaranta {\it et al.}, Phys. Rev. Lett. \textbf{9}, 226 (1962);
Y.-S. Tsai, Rev. Mod. Phys. \textbf{46}, 815 (1974).

\bibitem{BeBa} % free and bound-free muon production in Z-Z collision
C. A. Bertulani and G. Baur, Phys. Rep. {\bf 163}, 299 (1988);
J. Eichler, {\it ibid.} {\bf 193}, 165 (1990).

\bibitem{two-photon} We point out that the notion of ``two-photon pair production'' is also used in the context of heavy-ion collisions, where it refers to two {\it virtual} photons that are absorbed from the colliding Coulomb fields.

\bibitem{LL} V. B. Berestetskii, E. M. Lifshitz, and L. P. Pitaevskii,
{\it Relativistic Quantum Theory} (Pergamon, Oxford, 1971).

\bibitem{Povh} B. Povh, K. Rith, C. Scholz, and F. Zetsche, {\it Particles and Nuclei} (Springer, Berlin, 2006).

\bibitem{Thomas} Nuclear photoexcitation is unlikely as the doppler-shifted photon energy lies far above the typical energy $\lesssim 10$\,MeV of nuclear resonances [see, e.g., T. J. B\"{u}rvenich, J. Evers, and C. H. Keitel, Phys. Rev. Lett. \textbf{96}, 142501 (2006)].

\bibitem{futureLCLS} M. Cornacchia {\it et al.}, J. Synchrotron Radiat. \textbf{11}, 227 (2004); A. A. Zholents and W. M. Fawley, Phys. Rev. Lett. \textbf{92}, 224801 (2004); C. G. Schroer and B. Lengeler, {\it ibid.} \textbf{94}, 054802 (2005); A. Jarre {\it et al.}, {\it ibid.} \textbf{94}, 074801 (2005); D. D. Ryutov, Rev. Sci. Instrum. \textbf{76}, 023113 (2005); V. M. Malkin and N. J. Fisch, Phys. Rev. Lett. \textbf{99}, 205001 (2007).

\bibitem{radii} I. Angeli, At. Data Nucl. Data Tables \textbf{87}, 185 (2004); 
P. J. Mohr and B. N. Taylor, Rev. Mod. Phys. \textbf{77}, 1 (2005).

\bibitem{PRL} C. M\"uller, A. B. Voitkiv, and N. Gr\"un, Phys. Rev. Lett. {\bf 91}, 223601 (2003).

\bibitem{Sorensen} % bound-free muon production in \gamma-Z and Z-Z collision
A. Belkacem and A. H. S{\o}rensen, Phys. Rev. A {\bf 57}, 3646 (1998);
% free muon production in \gamma-Z collision
D. Ivanov and K. Melnikov, Phys. Rev. D {\bf 57}, 4025 (1998).

%\bibitem{RHIexp} D. Antreasyan {\it et al.}, CERN report EP/80-82 (1980).
%F. Vanucci {\it et al.}, CERN-EP/80-82 (http://cdsweb.cern.ch);

\bibitem{Mombi} % bound-free muon production in U-U collision
K. Momberger, N. Gr\"un, W. Scheid, U. Becker, and G. Soff, J. Phys. B {\bf 20}, L281 (1987).

\bibitem{exclusive}
A. Abulencia {\it et al.}, Phys. Rev. Lett. \textbf{98}, 112001 (2007); K. Hencken, E. A. Kuraev, and V. G. Serbo, Phys. Rev. C {\bf 75}, 034903 (2007).

\end{thebibliography}
\end{document}